\documentclass[letterpaper,twocolumn,aps]{revtex4}

\usepackage{graphicx}

\begin{document}

\title{Universal quasiparticle decoherence in hole- and electron-doped high-$T_c$ cuprates }

\author{Z.-H. Pan,$^{1}$ P. Richard,$^{1}$ A.V. Fedorov,$^{2}$ T. Kondo,$^{3}$ T. Takeuchi,$^{3}$ 
S.L.~Li,$^{4}$ Pengcheng Dai,$^{4,5}$ G.D. Gu,$^{6}$ W. Ku,$^{6}$ Z. Wang,$^{1}$ and H. Ding$^{1}$}

\affiliation{
(1) Department of Physics, Boston College, Chestnut Hill, MA 02467 \\
(2) Advanced Light Source, Lawrence Berkeley National Laboratory, Berkeley, CA 94720\\
(3) Research center for advanced waste and emission management, Nagoya University, Japan\\
(4) Department of Physics and Astronomy, University of Tennessee, Knoxville, TN 37996\\
(5) Neutron Scattering Sciences Division, Oak Ridge National Laboratory, Oak Ridge, TN 37831\\
(6) Condensed Matter and Materials Science Department, Brookhaven National Laboratory, Upton, New York 11973
}
\begin{abstract}
\noindent 
We use angle-resolved photoemission to unravel the quasiparticle decoherence process in the high-$T_c$ cuprates.
The coherent band is highly renormalized, and the incoherent part manifests itself as a nearly vertical ``dive'' in the $E$-$k$ 
intensity plot that approaches the bare band bottom. We find that the coherence-incoherence crossover energies in the hole- and 
electron-doped cuprates are quite different, but scale to their corresponding bare bandwidth. This rules out antiferromagnetic fluctuations 
as the main source for decoherence. We also observe the coherent band bottom at the zone center, whose intensity is strongly 
suppressed by the decoherence process. Consequently, the coherent band dispersion for both hole- and electron-doped cuprates is obtained, 
and is qualitatively consistent with the framework of Gutzwiller projection.
\noindent 
%\pacs{71.27.+a, 71.18.+y, 74.25.Jb, 74.70.-b}

\end{abstract}
\maketitle
Understanding the electronic structure and properties in strongly correlated systems, 
in particular in the high--$T_c$ cuprates, has been a main focus in condensed matter physics over the past two decades.
Unlike in simple metals and insulators, the presence of strong correlation makes the predictions of band calculations
such as the LDA unreliable. Much of the theoretical understanding, based upon studies of Hubbard-like models,
is the existence of strongly renormalized coherent quasiparticle excitations of a much reduced bandwidth $\textit{\`a la}$ Brinkman-Rice
\cite{brinkman_rice} and Gutzwiller wavefunctions \cite{gutzwiller}, and large incoherent background
of the Mott-Hubbard \cite{hubbard} type that extends to the bare band edge \cite{wangbangkotliar}.
Both features have been ubiquitously observed in the cuprates, yet the precise
description of the one-particle spectral function over the range of bare bandwidth provided by LDA remained incomplete
due to the complexity of this many-body problem. The correlation-induced thermodynamic mass
renormalization in the cuprate is $\sim3$, which can be directly extracted from the renormalized Fermi velocity
determined from angle-resolved photoelectron spectroscopy (ARPES) in the prototype Bi$_2$Sr$_2$CaCu$_2$O$_{8+\delta}$. 
Many ARPES measurements have been performed to determine the band dispersion in the high-$T_c$ cuprates. Remarkably,
the band dispersion for the quasiparticle excitations can only be traced up to
$\sim350$ meV, above which the band seems to disappear and the anticipated band bottom at the
$\Gamma$ (0,0) point has not been identified \cite{Norman_95}. 
A recent ARPES on the undoped cuprate Ca$_2$CuO$_2$Cl$_2$ also found that the renormalized band is truncated around 350 meV, 
and the incoherent part at high energy seems to follow the bare band dispersion predicted by LDA \cite{Ronning_05}. 
It was suggested that the cause of this truncation is the antiferromagnetic fluctuations, with a characteristic energy scale of 2$J$, 
where $J\sim120-160$ meV is the superexchange coupling of the Cu-O square lattice.

In this Letter, we report a systematic APRES study on the complete band dispersion of various cuprates, including hole-doped
Pb$_x$Bi$_{2-x}$Sr$_2$CuO$_{6+\delta}$ and electron-doped Pr$_{1-x}$LaCe$_x$CuO$_4$. 
The most important of our findings is that the truncation energy scale of the coherence-incoherence crossover is $\textit{not}$ 
fixed around 350 meV, or $\sim$ 2$J$, instead it is determined by and scales with the bare bandwidth. In the electron-doped cuprates, this crossover 
along $\Gamma-X$ occurs around $\sim$ 600 meV, much larger than the value (350 meV) in the hole-doped ones. At binding 
energies above the crossover, the incoherent part of the spectrum takes the form of a nearly vertical dispersion around a fixed crystal
momentum $k$, and approaches the bare band bottom predicted by LDA.  In addition, we observe for the first time the bottom of the renormalized 
band whose intensity is strongly suppressed.  The complete determination of the coherent part of the occupied Cu$3d_{x^2-y^2}$ 
band enables us to provide the tight-binding parameters and compare to the more realistic quasiparticle dispersion calculated
from models with strong local correlation. We find that the renormalized dispersion obtained from the Gutzwiller projected wavefunction approach
to the $t-J$ like models is a promising candidate for the observed low energy quasiparticle band.

High quality single crystals of cuprates Pb$_x$Bi$_{2-x}$Sr$_2$CuO$_{6+\delta}$ (Pb-Bi2201), Bi$_2$Sr$_2$CaCu$_2$O$_{8+\delta}$ 
(Bi2212), and Pr$_{1-x}$LaCe$_x$CuO$_4$ (PLCCO) were prepared by the the traveling solvent floating zone method, 
and some were annealed subsequently. 
ARPES experiments were performed at the Synchrotron Radiation Center, WI, and the 
Advanced Light Source, CA. High-resolution undulator beamlines and Scienta analyzers with a capability of multi-angle 
detection have been used. The energies of photons were carefully chosen in order to enhance certain spectral features. The energy resolution 
is $\sim$ 10 - 30 meV, and the momentum resolution $\sim0.02$ \AA$^{-1}$. 
All the samples were cleaved and measured \emph{in situ} in a vacuum better than $8\times10^{-11}$ $Torr$ at low temperatures 
(14 - 40 K) on a flat (001) surface, and all the spectra shown below have been reproduced on multiple samples.

%========================== FIG 1  ===============================
\begin{figure}[{here}]
\includegraphics[ width = 8cm]{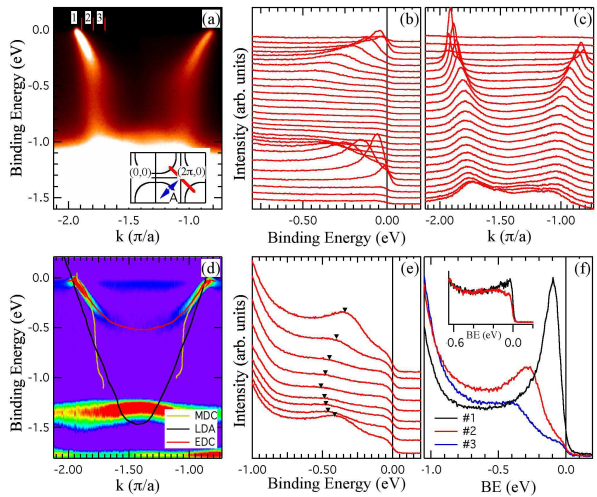}
\vspace{-10pt}
\caption{Dispersion of coherent and incoherent bands along  $\Gamma-X$ in Pb-Bi2201 
measured at 20K using 57-eV $s$-polarized photons.
(a) - (d) Plots of $E$-$k$ intensity, EDCs, MDCs, and the second derivative intensity, respectively. 
The inset in panel (a) displays the measurement locations in BZ. 
In panel (d), three extra lines of dispersion extracted from EDCs (red), MDCs (yellow), and LDA calculation (black) are also superimposed 
for the comparison purpose. 
(e) Magnified plot for EDCs near $\Gamma$. 
(f) Comparison of three EDCs at the $k$-locations labeled as $\#$1 to $\#$3 in Fig.~1a. The inset shows two EDCs of Bi2212 at the similar
$k$-locations as $\#$1 and $\#$2, but using $p$-polarized photons.
}
\label{VB}
%\vspace{-17pt}
\end{figure}
%=======================================================================

We start with a set of spectra along $\Gamma-X$ on a hole-doped Pb$_x$Bi$_{2-x}$Sr$_2$CuO$_{6+\delta}$ (overdoped $T_c \sim$ 7K), 
as shown in Fig.~1. The reason we choose this material is that the Pb substitutions remove the superlattice modulation in Bi-O plane, 
which often complicates ARPES spectra \cite{Ding_PRL96}. The spectra were taken along $\Gamma-X$ in the
second Brillouin zone (BZ) using 57-eV $s$-polarized ($\vec{A} \parallel \Gamma$$X$) photons to enhance various features at high binding energy ($>$ 350 meV). 
In Fig.~1a, one can easily follow the dispersive band at low energy ($<$ 350 meV). This band is the well-know Zhang-Rice singlet \cite{ZRS},
with the predominant Cu$3d_{x^2-y^2}$-O$2p_{x,y}$ antibonding orbital. While the Fermi vector $k_F$ is almost the same as the one predicted 
by LDA, as shown in Fig.~1d, its dispersion velocity $v_k$ ($\sim$ 2.1 eV\AA) is much smaller than the LDA value ($\sim$ 5.2 eV\AA) \cite{LDA_1},
consistent with previous ARPES results \cite{Dessau_93,Ding_PRL96,Valla_PRL00}. Note that the Fermi velocity $v_F$ ($\sim$ 1.6 eV\AA) at
$k_F$ is even smaller due to a further renormalization by the observed nodal kink at $\sim$ 70 meV  \cite{Valla_Science,Lonraza_Nature}, 
which is difficult to visualize at the large energy scale in Fig.~1. At higher binding energy ($>$ 350 meV), the spectrum becomes
ill-defined. While the intensity plot (Fig.~1a) seems to indicate that the spectra abruptly ``dive'' almost vertically from 350 meV to at least 1 eV, 
this diving behavior does not expressed itself as a peak in the energy distribution curves (EDCs) (Fig.~1b). Instead,
an enhancement in the EDC background is observed at the $k$-location of the dive around ($\pi/4$, $\pi/4$) 
and equivalent $k$ points in other BZs, as shown in 
Fig.~1f where the EDC at the dive location ($\#$ 2, as marked in Fig.~1a) has a higher background than its neighboring EDCs
($e.g.$, $\#$ 1 and $\#$ 3). The diving behavior is reflected more clearly in the momentum 
distribution curves (MDCs), as shown in Figs.~1d where the MDCs seem to maintain their peak shape. 
Since the dive completely loses the peak (or pole) structure in energy, strictly it is no longer a band. Nevertheless, it is likely the incoherent part of 
the Cu$3d_{x^2-y^2}$-O$2p_{x,y}$ (or ZRS) spectrum, since it maintains the $d_{x^2-y^2}$ symmetry. We have verified this symmetry, as shown in
the inset of Fig.~1f, where both the coherent band and the dive are suppressed by $p$-polarized light ($\vec{A} \perp \Gamma$$X$). This is due to 
a well-known ARPES selection rule  \cite{Norman_selection}. In addition, we have also observed that the intensity ratio between these two features is
roughly a constant when we change $s$ and $p$ polarization components, supporting that the dive is the incoherent part of the band.

So far we have shown that the coherent ZRS band disperses to a certain energy ($\sim$ 350 meV) and then abruptly switches to the incoherent part 
at the higher energies. However, if we take a closer look at the EDCs in the vicinity of $\Gamma$, we observe the 
smooth continuation of the coherent band, which reaches the bottom around 0.5 eV at $\Gamma$, as shown in Fig.~1e. 
This is the long-sought-after renormalized band bottom, and the reason we can observe it for the first time is due to several combined factors 
such as the 
superlattice free sample, a proper photon energy, and the second BZ, all of which enhance the intensity of the high-energy features.
We note that the coherent band starts to lose its intensity at the same energy where the incoherent diving pattern begins to form, 
indicating a weight transfer between the coherent and incoherent parts. The bottom of the coherent band can
be also visualized, as shown in Fig.~1d, from the second derivative of the intensity with respect to energy which enhances broad horizontal features.
In Fig.~1d, we compare the dispersion of the coherent band with the calculated one from LDA, along with the dispersion extracted from MDCs.
It is clear that the coherent band, with a well-defined parabolic shape, is highly renormalized, and the vertical feature is likely the incoherent part, which
approaches the bottom of the bare band.

%========================== FIG 2  ===============================
\begin{figure}[{here}]

\includegraphics[ width = 8cm]{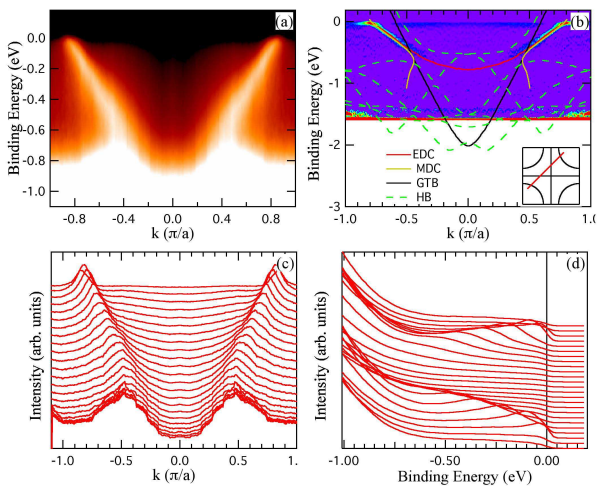}
\vspace{-10pt}
\caption{
Dispersion of the coherent and incoherent bands near $\Gamma-X$ in PLCCO measured at 40 K using
22-eV photons.
(a) - (d) Plots of $E$-$k$ intensity, second derivative intensity, EDCs, and MDCs, respectively. 
The inset in panel (b) displays the measurement locations in BZ.
The superimposed curves in panel (b) are the extracted EDC positions (red), fitted MDC positions (yellow), non-hybridized LDA band (solid
black), and hybridized LDA bands (dashed green).
}
\label{VB}
%\vspace{-17pt}
\end{figure}
%=======================================================================

We have observed very similar behaviors of the coherence-incoherence crossover in the bilayer system Bi$_2$Sr$_2$CaCu$_2$O$_{8+\delta}$,
except the superlattice in this material makes it difficult to observe the bottom formation of the coherent band at $\Gamma$. Like in Pb-Bi2201,
the incoherent part in Bi2212 deviates from the coherent part around 350 meV, and approaches almost vertically the bare band bottom around
1.5 eV. Since the 350 meV energy scale of the coherence-incoherence crossover has been attributed to the antiferromagnetic fluctuations 
whose characteristic energy scale 2$J$ has a similar value \cite{Ronning_05}, it is natural to check if a similar coherence-incoherence 
crossover exists in the electron-doped cuprates. We have searched for this crossover on various electron-doped cuprates, 
and the main results are presented in the following two figures.

We first show, in Fig.~2, the dispersion of the electron-doped cuprate Pr$_{0.88}$LaCe$_{0.12}$CuO$_4$ ($T_c \sim$ 23 K) near $\Gamma-X$
using 22-eV $p$-polarized photons. 
Since the band intensity exactly along $\Gamma-X$ is highly suppressed due to the selection rule mentioned above, we choose to display 
the dispersion along the parallel direction slightly away from the $\Gamma-X$ direction, as indicted in the inset of Fig.~2b.
A quick examination of the plots of Fig.~2 reveals a major difference to the hole-doped materials: the coherent band in PLCCO extends to a much higher
binding energy. The separation of the incoherent part occurs around 0.6-0.7 eV, which also forms a diving pattern at higher binding energy, 
as seen in both the intensity plot in Fig.~2a and the MDCs plot in Fig.~2c. We note that the diving pattern appears to be shorter than the one in the 
hole-doped cuprates. We believe that this is due to the hybridization between the Cu$3d_{x^2-y^2}$ band and some other bands, as predicted
by LDA calculations \cite{NCCO_LDA} and shown in Fig.~2b (dashed green curves). The bare Cu$3d_{x^2-y^2}$ band, while not mixing with those
bands, reaches the bottom around 2.1 eV, as indicted in Fig.~2b (solid black curves) \cite{NCCO_GTB}. In comparison, the coherent part has its 
band bottom around 0.8 eV, indicating a mass renormalization of 2.5, similar as in the hole-doped case.

%====================   FIG 3   =====================
\begin{figure}[{here}]
\includegraphics[width = 8cm]{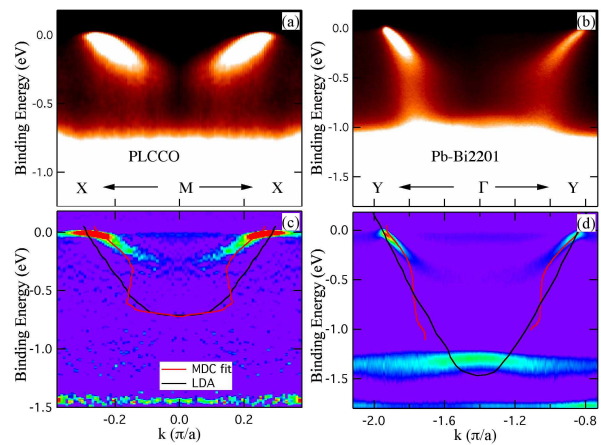}
\vspace{-10pt}
\caption{
Comparison of band dispersion between $M-X$ in PLCCO using 22-eV photons and $\Gamma-X$ in Pb-Bi2201 using 57-eV photons. 
(a) - (b) $E$-$k$ intensity plots for $M-X$ in PLCCO and
$\Gamma-X$ in Bi2201, respectively. (c) - (d) Corresponding second derivative plots for panels (a) and (b), respectively. 
MDC dispersion (red lines) and bare band from LDA (black lines) are superimposed to the plots in panels (c) and (d).
}
\label{Crossing}
%\vspace{-17pt}
\end{figure}
%==================================================

The different energy scale of the coherence-incoherence crossover in the electron-doped cuprates, as observed in PLCCO and confirmed in other 
electron-doped materials, such as Pr$_{2-x}$Ce$_x$CuO$_4$ and Nd$_{2-x}$Ce$_x$CuO$_4$, argues strongly against the antiferromagnetic 
scenario. Instead, the same mass renormalization ratio in both cases suggests that this energy scale may be related to the bare bandwidth. 
This is also true in the electron-doped material along another high-symmetry direction, $M-X$, as can be seen in Fig.~3.
It is well known that the van-Hove saddle point shifts to a much higher binding energy ($\sim$ 0.4 eV) in the electron-doped cuprates 
\cite{King_NCCO,Takahashi_NCCO}. Over the wide energy range, the  Cu$3d_{x^2-y^2}$ band dispersion along $M-X$ in PLCCO 
has many similarities to the band along $\Gamma-X$ in Bi2201, as shown in Fig.~3.  The coherent part along $M-X$ in PLCCO, while being 
quite broad due to the possible stronger interactions near the antinode, extends to an energy scale ($\sim$ 0.25 eV) when the incoherent 
part takes a dive. The bottom of the coherent part at $M$ is estimated to be $\sim$ 0.3 eV, and the bare band position at $M$ is calculated 
by LDA to be $\sim$ 0.7 eV. For comparison, we draw both intensity plot and second derivative plot of Pb-Bi2201 in the left panels of Fig.~3 in a slightly 
reduced energy scale. We also notice that there is a small kink at $\sim$ 70 meV along $M-X$ in PLCCO, 
which resembles the well-know and much-debated nodal kink in the holed-doped cuprates \cite{Valla_Science,Lonraza_Nature}. 
We caution that the origin of the antinodal kink in PLCCO may not be the same as the antinodal one in the hole-doped case, and
call for more systematic studies.

We have measured band dispersion along many directions in the BZ for various hole- and electron-doped cuprates. In Fig.~4, 
we summarize our main results as a comparison between the coherent band dispersion and the bare band dispersion predicted by LDA along 
several high-symmetry directions, for both Pb-Bi2201 ($T_c \sim$ 7 K) and PLCCO ($T_c \sim$ 23 K) samples. We also use the effective 
tight-binding band to fit the coherent dispersion, 
using the standard formula as in previous work \cite{Norman_95}. Since ARPES only measures the occupied side, we adapt the previous method
\cite{Norman_95} by choosing the unoccupied band top at $X$ ($\pi, \pi$) in such a way that it maintains the same band renormalization ratio 
($\sim$ 2.5) as the occupied side. We use the six free parameters ($t_0$ to $t_5$) with $t_0$ being the chemical potential, $t_1$ the 
nearest neighboring hopping term, and $t_2$ to $t_5$ the higher order hopping terms. The numeric values of these parameters are also listed 
in Fig.~4. The large difference of $t_0$ ($\sim$ 0.4 eV) between the two systems indicates a large chemical shift from the hole doped side to 
the electron doped side, which is likely the main cause of the downshift of ($\sim$ 0.4 eV) of the van Hove singularity in the electron-doped 
cuprates.

%=================FIG 4 ================================================
\begin{figure}[{top}]
\includegraphics[width = 8cm]{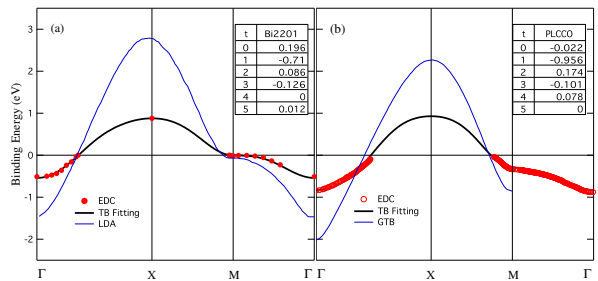}
%\vspace{-10pt}
\caption{
Summary of coherent band dispersion along three principle directions ($\Gamma$-$X$, $X$-$M$, and $M$-$\Gamma$) for 
hole- and electron-doped cuprates. (a) - (b) Measured coherent band position (red dots), tight-binding fit (black solid line), 
and LDA band dispersion (blue dashed line) in Pb-Bi2201 and
PLCCO, respectively.
The inserted tables are the obtained fitting parameters.
}
\label{FS}
%\vspace{-17pt}
\end{figure}
%=====================================================================

In summary, we have determined by ARPES the complete low energy quasiparticle dispersion and elucidated
its unusual evolution to the high energy incoherent background in both hole- and electron-doped cuprates. The reduction
of the bandwidth from its bare value is most likely the result of strong local correlations
that frustrate the kinetic energy. This is overall consistent with the Gutzwiller projected wavefunction approach to simple
models of doped Mott insulators \cite{AtoZ_paper}. Detailed comparisons would require measurement of the doping dependence
of the renormalized bandwidth, which is more difficult to determine due to the doping dependent shift 
of the chemical potential. More systematic studies are needed to clarify this issue.
Most surprisingly, we find that the energy scale associated
with the coherence-incoherence crossover is determined by a fraction of the bare band bottom energy and is in general
different from the antiferromagnetic exchange energy 2$J$, ruling out the latter as the main cause of quasiparticle decoherence
at {\it high} binding energies. Moreover, it appears more universal that the incoherent spectrum beyond the decoherence
energy takes a vertical dive with a nearly fixed $k\sim(\pi/4,\pi/4)$, approaching the bottom of the bare band in what seems
to be the ``cheapest'' way for the renormalized quasiparticles to return to their bare form. The origin of these
unexpected behaviors is largely unknown and demands more understanding of the interplay between the coherent quasiparticle 
and collective excitations and the dominant incoherent processes in the spectral function, which has been one of the central 
challenges in the physics of strong correlations.

We thank P.W. Anderson, D.H. Lee, M.R. Norman, F.C. Zhang for valuable discussions and suggestions.
This work is supported by NSF DMR-0353108, DOE DE-FG02-99ER45747, DE-FG02-05ER46202, and DE-AC02-98CH10886. 
ZHP is also supported by ALS Doctoral Fellowship in Residence. This work is based upon research conducted at 
the Synchrotron Radiation Center supported by NSF DMR-0084402, and at the Advanced Light Source supported by DOE DE-AC03-76SF00098. 
ORNL is supported by DOE under contract No. DE-AC05-00OR22725 with UT/Battelle, LLC.

\textit{Note added}: During the preparation of this manuscript, we became aware of three preprints reporting independently ARPES results of band 
dispersion over a large energy scale on the hole-doped cuprates \cite{Lanzara_preprint,Feng_preprint,Valla_preprint}.

\end{document}